\documentclass[prl,twocolumn]{revtex4}
\usepackage{epsfig}
\usepackage{bm}

\begin{document}

\title{Hamiltonian properties of Madden-Julian Oscillation: waves and distributed chaos}

\author{A. Bershadskii}

\affiliation{
ICAR, P.O. Box 31155, Jerusalem 91000, Israel
}

\begin{abstract}
  
  The Madden-Julian Oscillation (MJO) has been studied using its daily RMM1, RMM2 and VPM1, VPM2 principal components' indices. A spectral analysis of the raw indices indicates presence of a noise presumably contributed by the random high-frequency waves. A soft simple filter - 3 day running average, has been applied to filter out this high-frequency noise (the filter has been chosen in accordance with the observationally known properties of these waves). Power spectrum of the smoothed indices exhibits a stretched exponential decay $E(f) \propto \exp-(f/f_0)^{1/2}$ characteristic to the Hamiltonian distributed chaos with spontaneously broken time translational symmetry (with time depending Hamiltonian and adiabatic invariance of the action). An impact of the MJO's Hamiltonian distributed chaos on the Asian-Australian Monsoons at the intraseasonal time scales has been briefly discussed.

\end{abstract}

\maketitle

\section{Introduction}

At present time three main atmospheric phenomena are known to determine the tropic atmosphere and provide a crucial energy supply to the global atmospheric circulation from the daily to multi-annual time scales: the Asian-Australian monsoons, the El Ni\~no-Southern Oscillation (ENSO) and  the Madden-Julian Oscillation (MJO). In this note the MJO will be studied as a Hamiltonian phenomenon from the daily to the intraseasonal time scales (see Refs. \cite{zha},\cite{ven2} for a comprehensive review of the MJO phenomenon). 

\begin{figure} \vspace{-0.6cm}\centering
\epsfig{width=.45\textwidth,file=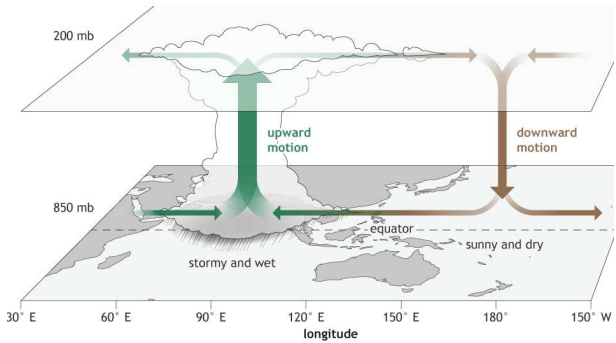} \vspace{-4.5cm}
\caption{A sketch of the upper-atmosphere and surface formation of the MJO (adapted from Ref. \cite{pic})}
\end{figure}
\begin{figure} \vspace{-0.5cm}\centering
\epsfig{width=.45\textwidth,file=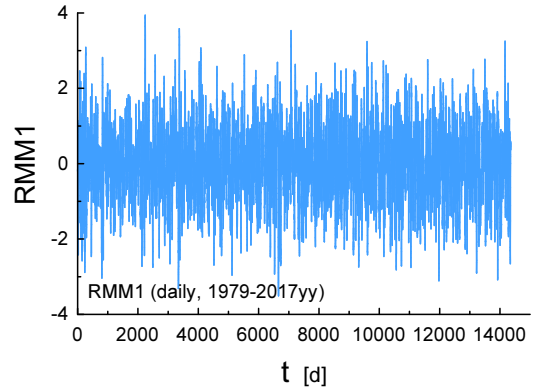} \vspace{-4cm}
\caption{The RMM1 daily time series. The data were taken from Ref. \cite{knmi}.}
\end{figure}
  Figure 1 shows a sketch of the upper-atmosphere and surface formation of the MJO (the drawing by Fiona Martin has been adapted from the Ref. \cite{pic}). In this crucial for the MJO stage thunderstorm cloud (the enhanced convective phase) is centered over the Indian Ocean while over the west-central Pacific Ocean the atmosphere is dominated by the suppressed convective phase. This formation traverse the globe in eastward direction (it mainly  manifests itself in
the Eastern Hemisphere) and returns to the initial stage with a period about 50 day on average (actually 30 to 60 days, see also below). Therefore more than one MJO events can occur within a season (intraseasonal climate variability Ref. \cite{zha} an references therein). Rather roughly it can be considered as global-scale coupled patterns in baroclinic circulation disturbances and deep convection  \cite{zha},\cite{ven2}. 

  The MJO plays a significant role in modulation of tropical cyclone activity \cite{ven1},\cite{bw}, of the monsoons \cite{zha},\cite{wh} and is intimately related to the ENSO \cite{lau},\cite{hwz}. It also has global teleconnections impacting extratropic climate and weather \cite{stan}-\cite{sh}.  
\begin{figure} \vspace{-0.3cm}\centering
\epsfig{width=.45\textwidth,file=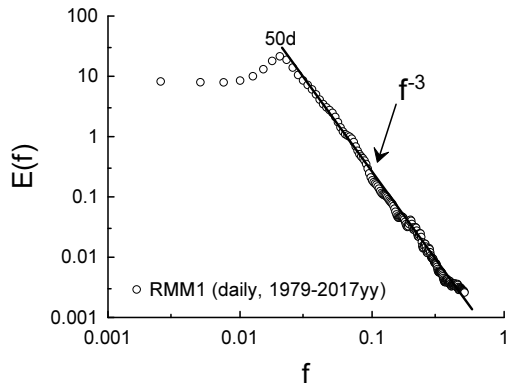} \vspace{-4.5cm}
\caption{Power spectrum for the time series shown in the Fig.2 (in the log-log scales) \cite{knmi}.}
\end{figure}
\begin{figure} \vspace{-0.6cm}\centering
\epsfig{width=.45\textwidth,file=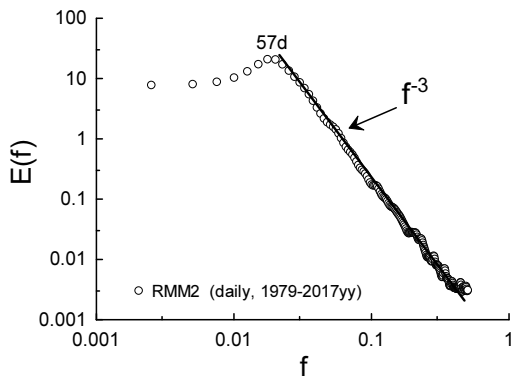} \vspace{-4.3cm}
\caption{As in Fig. 3 but for the RMM2 daily index \cite{knmi}.}
\end{figure}
\begin{figure} \vspace{-0.4cm}\centering
\epsfig{width=.45\textwidth,file=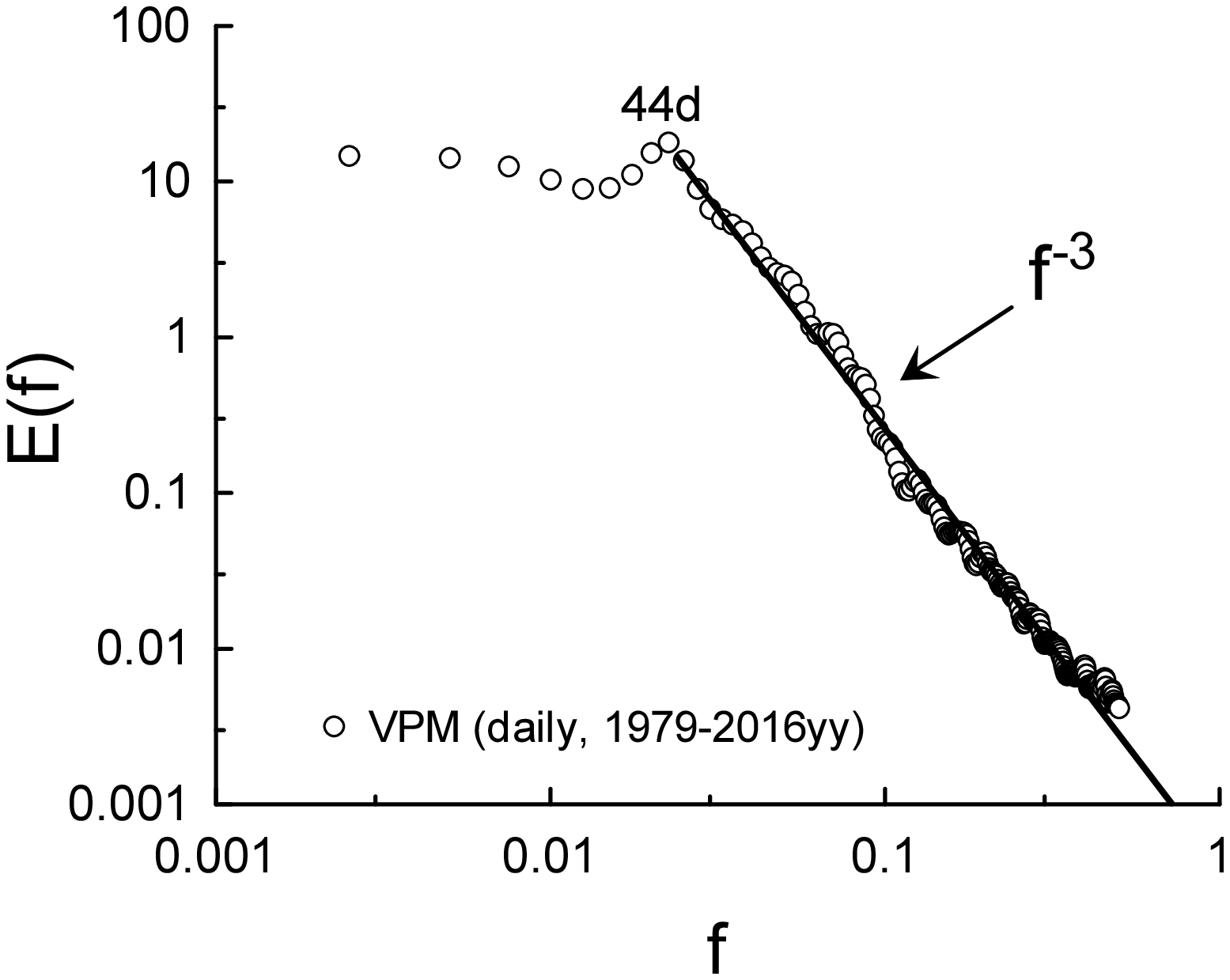} \vspace{-4.4cm}
\caption{As in Fig. 3 but for the VPM1 daily index \cite{noaa1}.}
\end{figure}
\begin{figure} \vspace{-0.4cm}\centering
\epsfig{width=.45\textwidth,file=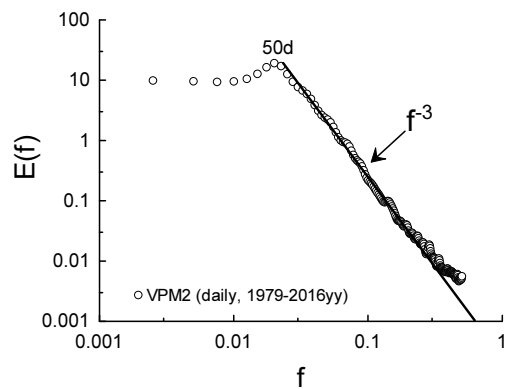} \vspace{-4.35cm}
\caption{As in Fig. 5 but for the VPM2 daily index \cite{noaa1}.}
\end{figure}
\section{MJO daily indices}
 The authors of the Ref. \cite{wh} constructed a daily index (actually two indices - RMM1 and RMM2) for monitoring the MJO. These indices are now widely used for the monitoring and predicting purposes. They are seasonally independent and based on a pair of EOFs (empirical orthogonal functions) of the near-equatorially averaged combined fields: 200-hPa zonal wind, 850-hPa zonal wind (see the Fig. 1), and satellite-observed OLR (outgoing long-wave radiation) data. The two daily time series - RMM1 and RMM2 (representing two principal components) were obtained by 
projection of the observed data onto the EOFs. The authors of the Ref. \cite{wh} believed that this projection by itself (without using a conventional time filtering, see also below) can serve as an sufficient filter for the MJO. The intraseasonal time scales, characteristic for the MJO only, were separated by removing of the components of interannual variability and annual cycle. 

   The two complimentary daily indices - VPM1 and VPM2, were suggested in the more recent Ref. \cite{ven2}. They were constructed like the RMM1 and RMM2 indices but a meridionally averaged 200-hPa velocity potential was used instead of the satellite-observed OLR. And again the conventional time filtering was not applied to the daily time series. \\
   
   Figure 2 shows the daily RMM1 time series for 1979-2017yy period (the data were taken from the Ref. \cite{knmi}). This is a statistically stationary time series. Figure 3 shows (in the log-log scales) corresponding power spectrum computed by the maximum entropy method, with an optimal resolution for short time series \cite{oh}. The straight line in the Fig. 3 indicates a scale-invariant (scaling) power law decay of the spectrum:
$$
E(f) \propto f^{-3} \eqno{(2)}
$$
\begin{figure} \vspace{-1.64cm}\centering
\epsfig{width=.45\textwidth,file=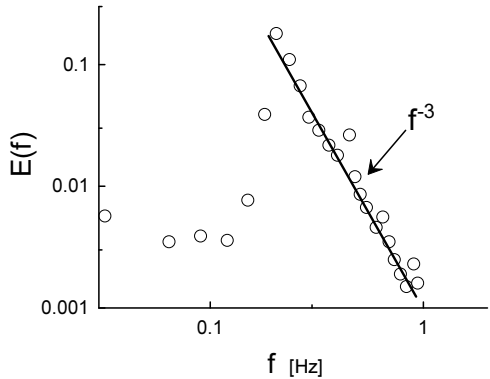} \vspace{-3.85cm}
\caption{A shallow sea wind wave energy spectrum in the log-log scales.}
\end{figure}
  Figure 4 shows analogous spectrum for the daily RMM2 index. Figures 5 and 6 show analogous power spectra for the daily VPM1 and VPM2 indices respectively (the data were taken from the Ref. \cite{noaa1}). The straight line in these figures indicates the scaling spectrum Eq. (2). 
  
  The red noise, for instance, has scaling spectrum $E(f) \propto f^{-2}$. An example of a more steep and relevant scaling decay of the power spectra is shown in Figure 7. In this figure results of shallow sea measurements for the wind wave energy spectrum are shown in the log-log scales. The spectral data were taken from the Ref. \cite{liu1}. The measurements were made at the surf zone in 2.8m water depth. The straight line in the Fig. 7 indicates the scaling power law decay of the spectrum - Eq. (2). The scaling decay observed in the Figs. 3-6 can be explained as a result of a noise produced by the relatively high-frequency random waves, common for the tropics (cf. the Ref. \cite{tk} and references therein, and see above about the omitted filtering of the daily indices). As it is mentioned in the Ref. \cite{tk} the horizontal dynamics of the waves resembles that for shallow-water inertial-gravity waves (cf. Fig. 7). The corresponding period of the high-frequency random waves is usually less than 3 days \cite{tk}. Therefore, let us apply a soft simple filter - 3 day running average, to the daily time series RMM1, RMM2, VPM1 and VPM2 in order to eliminate this noise.  
\begin{figure} \vspace{-1.15cm}\centering
\epsfig{width=.45\textwidth,file=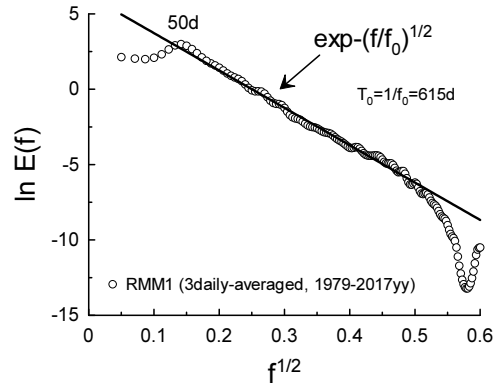} \vspace{-4.3cm}
\caption{Power spectrum of the daily RMM1 index filtered by the 3 day running average.}
\end{figure}
\begin{figure} \vspace{-0.2cm}\centering
\epsfig{width=.45\textwidth,file=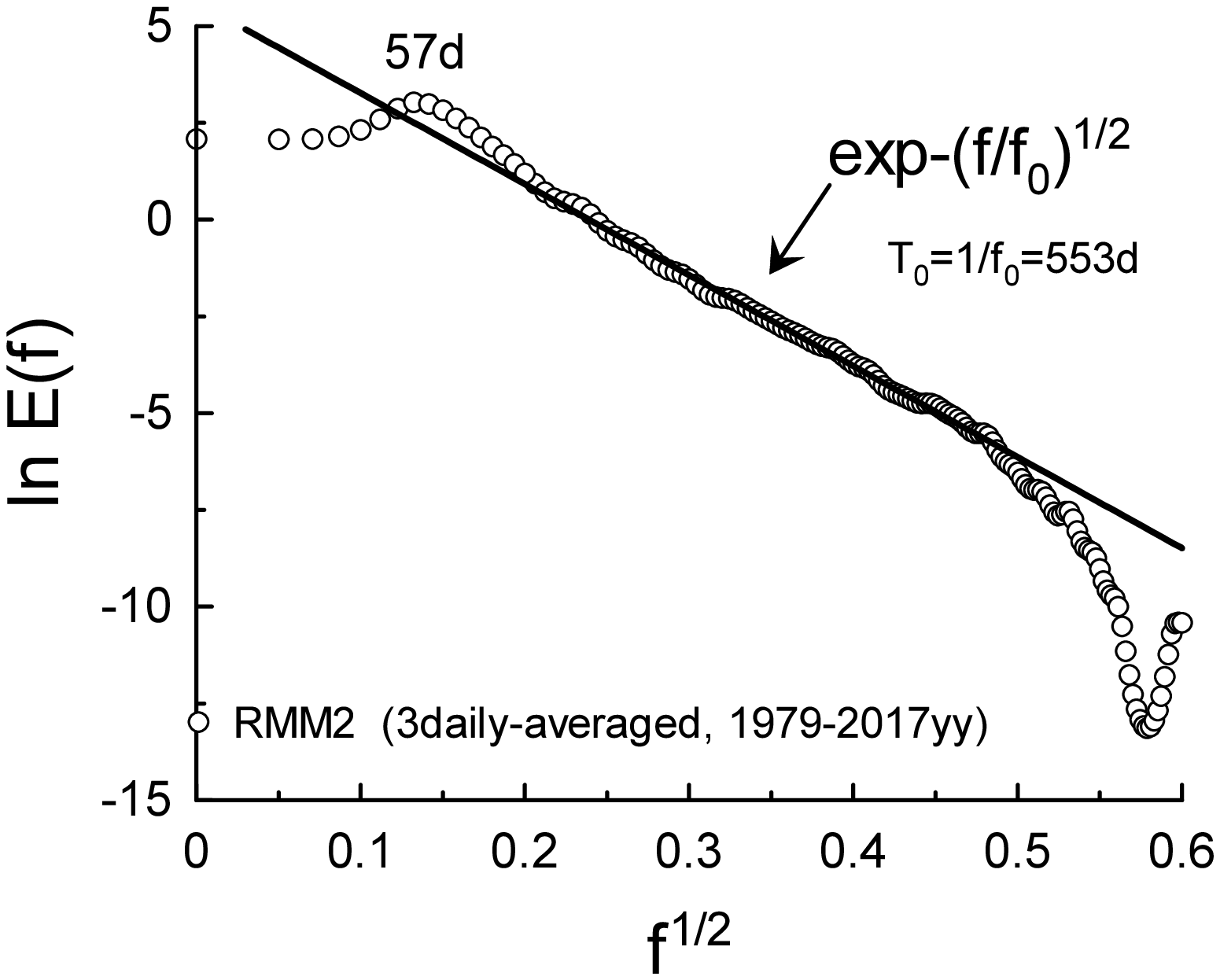} \vspace{-4.3cm}
\caption{As in Fig. 8 but for the daily RMM2 index filtered by the 3 day running average.}
\end{figure}
\begin{figure} \vspace{-0.7cm}\centering
\epsfig{width=.45\textwidth,file=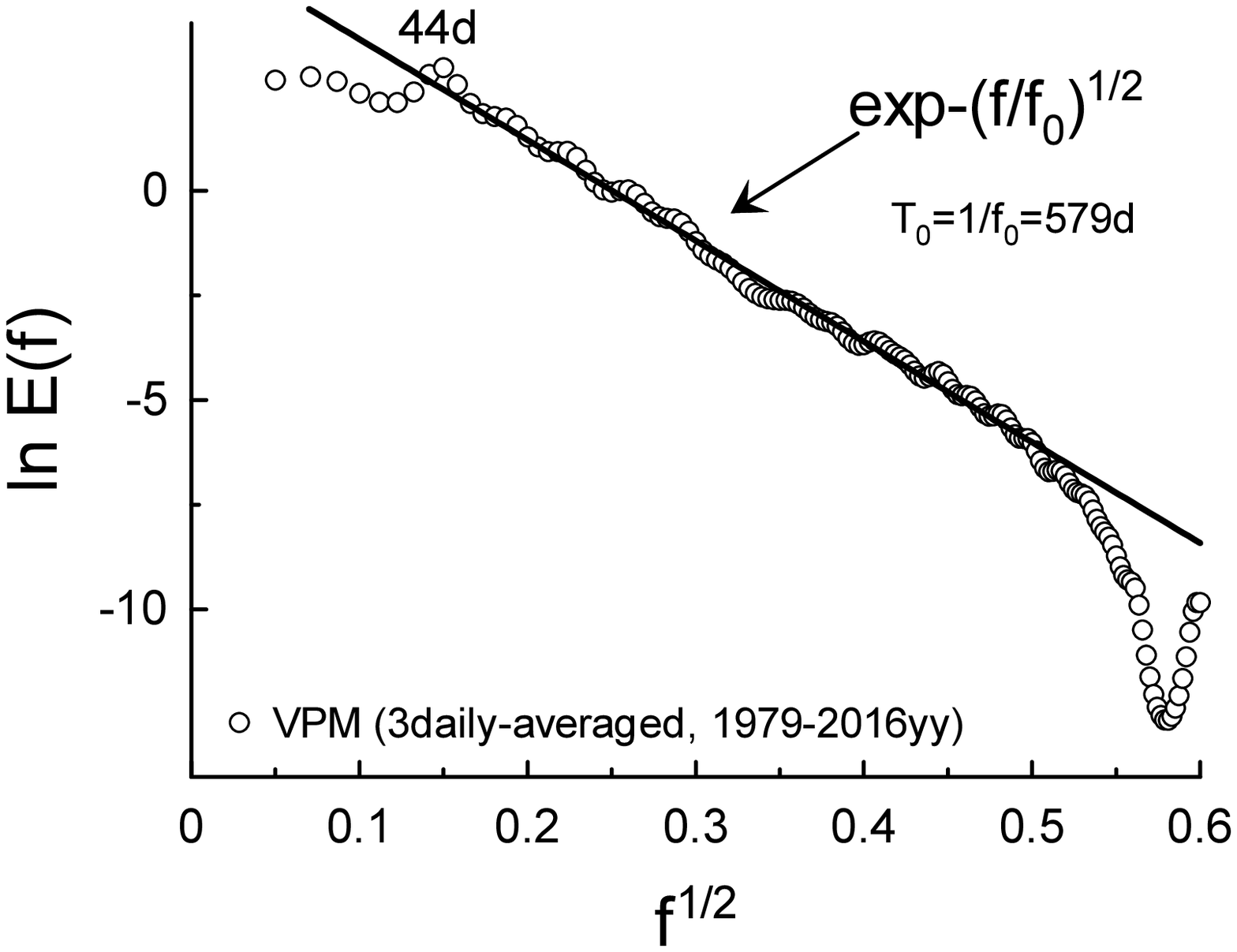} \vspace{-4.55cm}
\caption{Power spectrum of the daily VPM1 index filtered by the 3 day running average.}
\end{figure}
\begin{figure} \vspace{-0.5cm}\centering
\epsfig{width=.45\textwidth,file=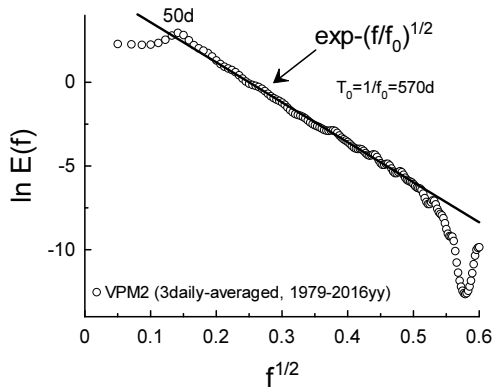} \vspace{-4.35cm}
\caption{As in Fig. 10 but for the daily VPM2 index filtered by the 3 day running average}
\end{figure}

\section{Hamiltonian distributed chaos}

In order to analyse the filtered (smoothed) indices let us recall certain properties of the  Hamiltonian distributed chaos \cite{b1}. Since most of the leading theoretical models of atmosphere and ocean dynamics are Hamiltonian \cite{she}-\cite{gl2} the Hamiltonian approach can be rather relevant in this case as well (see also Ref. \cite{b3}). 

 The exponential power spectra 
$$
E(f) \propto \exp -(f/f_c)      \eqno{(3)}
$$
where $f_c = const$ is some characteristic frequency, can be considered as typical for chaotic dynamical systems \cite{oh},\cite{sig}-\cite{fm}. For Hamiltonian sub-class of the dynamical systems a weighted superposition of the exponentials  
$$
E(f) \propto \int_0^{\infty} P(f_c)~ \exp -(f/f_c)~ df_c   \eqno{(4)}
$$
replaces the simple spectrum Eq. (3). Here $P(f_c )$ is a weight (probability distribution) of $f_c$.

  The energy conservation law for the Hamiltonian systems is a consequence of the time translational symmetry (invariance) - the Noether's theorem \cite{she}. When the time translational symmetry  is spontaneously broken the adiabatic invariance of action $I$ replaces the energy conservation (the Hamiltonian becomes time dependent) \cite{suz}. For these systems the dimensional considerations can be used to determine a relationship of characteristic velocity $v_c$ with the characteristic frequency $f_c$:
$$    
v_c \propto I^{1/2} f_c^{1/2}  \eqno{(5)}
$$
Normal (Gaussian) distribution of the characteristic velocity results in the chi-squared ($\chi^{2}$) distribution of the characteristic frequency $f_c$ 
$$
P(f_c) \propto f_c^{-1/2} \exp-(f_c/4f_0)  \eqno{(6)}
$$
$f_0$ is a constant. 

   Substitution of the Eq. (6) into Eq. (4) results in
$$
E(f) \propto \exp-(f/f_0)^{1/2}  \eqno{(7)}
$$

  Figure 8 shows power spectrum of the daily RMM1 index filtered by the 3 day running average (cf. Fig. 3). The straight line (a best fit in the appropriately chosen scales) indicates the spectral decay corresponding to the Hamiltonian distributed chaos Eq. (7). Figure 9 shows analogous spectrum for the filtered RMM2 index. Figures 10 and 11 show analogous spectrum for the filtered VPM1 and VPM2 indices. One can see that the Hamiltonian distributed chaos dominates the spectral decay for all four cases.\\

 The fundamental period $\mathrm{T}_f = 50 \pm 7$ day of the Hamiltonian distributed chaos corresponds to the MJO return period (see Introduction). Comparing these results with those obtained in the Ref. \cite{b3} one can conclude that the MJO Hamiltonian dynamics pumps the Asian-Australian monsoons' distributed chaos at the intraseasonal scales (cf. Refs. \cite{ven2},\cite{mm},\cite{kul} and references therein).

\section{Acknowledgement}

I thank B. Galperin for stimulating discussion. I acknowledge using the data provided by the KNMI Climate Explorer, the Australian  Bureau of Meteorology and the NOAA/National Weather Service (USA).

\end{document}